\documentclass[aps,prb,preprint,groupedaddress,showpacs]{revtex4}
\usepackage{graphicx}
\usepackage{amssymb}
\usepackage{amsmath}
\usepackage{latexsym}
\usepackage{ulem}
\usepackage{color}
\usepackage{dcolumn}
\usepackage{subfigure}

\begin{document}

\title{Nanoscale inhomogeneities: A new path toward high Curie temperature ferromagnetism in diluted materials}

\author{Akash Chakraborty}
\affiliation{
Institut N\'eel, CNRS, D\'epartement MCBT, 25 avenue des Martyrs, B.P. 166, 38042 Grenoble Cedex 09, France
}
\author{Richard Bouzerar}
\affiliation{
Institut N\'eel, CNRS, D\'epartement MCBT, 25 avenue des Martyrs, B.P. 166, 38042 Grenoble Cedex 09, France
}
\affiliation{
European Synchrotron Radiation Facility, B.P. 220, F-38043 Grenoble Cedex, France
}
\author{Stefan Kettemann}
\affiliation{
Division of Advanced Materials Science, Pohang University of Science and Technology (POSTECH), Pohang 790-784, South Korea
}
\affiliation{
School of Engineering and Science, Jacobs University Bremen, Campus Ring 1, D-28759 Bremen, Germany
}
\author{Georges Bouzerar}
\email{georges.bouzerar@grenoble.cnrs.fr}
\affiliation{
Institut N\'eel, CNRS, D\'epartement MCBT, 25 avenue des Martyrs, B.P. 166, 38042 Grenoble Cedex 09, France
}
\affiliation{
School of Engineering and Science, Jacobs University Bremen, Campus Ring 1, D-28759 Bremen, Germany
}

\date{\today}

\begin{abstract}
Room temperature ferromagnetism has been one of the most sought after topics in today's emerging field of spintronics. It is strongly believed that defect- and inhomogeneity-
 free sample growth should be the optimal route for achieving room-temperature ferromagnetism and  huge efforts are made in order to grow samples as ``clean" as possible. However, 
until now, in the dilute regime it has been difficult to obtain Curie temperatures larger than that measured in well annealed samples of (Ga,Mn)As ($\sim$190 K for 12\% doping).
In the present work, we propose an innovative path to room temperature ferromagnetism in diluted magnetic semiconductors.  We theoretically show that even a very 
small concentration of nanoscale inhomogeneities can lead to a tremendous boost of the critical temperatures: up to a 1600\% increase compared to the homogeneous case. In 
addition to a very detailed analysis, we also give a plausible explanation for the wide variation of the critical temperatures observed in (Ga,Mn)N and provide a better 
understanding of the likely origin of very high Curie temperatures measured occasionally in some cases. The colossal increase of the ordering temperatures by nanoscale 
cluster inclusions should open up a new direction toward the synthesis of materials relevant for spintronic functionalities.
\end{abstract}

\pacs{75.50.Pp, 75.30.Kz, 75.40.-s}

\maketitle

\section{INTRODUCTION}
The hope of attaining ferromagnetic order at room temperature and above, has spurred a huge interest in the field 
of diluted magnetic semiconductors (DMSs)\cite{satormp,jungwirth,timm} and diluted magnetic oxides (DMOs)\cite{sato-yoshida,fukumura,chambers}. Extensive 
experimental as well as theoretical efforts have been made to predict high Curie temperatures (T$_C$) in these materials. Among the various materials widely studied, 
one of particular interest is (Ga,Mn)N, a wide-band-gap DMS. Different experimental results have reported T$_C$'s varying as widely as 10 to 
940 K\cite{overberg,theodoropoulou,reed,thaler,sonoda} with a typical Mn content between 7 and 9\%. However, recent theoretical studies, based on model calculations, 
have predicted a T$_C$ of 30 K in homogeneously diluted and uncompensated Ga$_{1-x}$Mn$_{x}$N for $x$=0.06\cite{richard1}, which is in good agreement with results 
obtained from \textit{ab initio} based studies \cite{georges1} combined with the self-consistent local random-phase approximation (SC-LRPA) method. On the other hand, using the same ab initio couplings, the 
Monte Carlo studies lead to T$_C$ of 35 K for $x$=0.06\cite{bergqvist,satoprb} in Ga$_{1-x}$Mn$_{x}$N. These theoretical calculations predict the highest reachable T$_C$ 
in homogeneously diluted Ga$_{1-x}$Mn$_{x}$N. Then how can we explain the very high Curie temperatures observed by some experimental groups? We will provide an answer in 
the following. 

From these observations crucial questions arise. How do we explain the huge fluctuations of the critical temperatures in these materials? Is there a systematic
 way to boost the critical temperatures beyond that expected in the homogeneous compounds (inhomogeneity free)? 
 After observation of ferromagnetic order in Mn-doped Germanium (T$_C$=116 K for $x$=0.035)\cite{park}, several experimental studies reported quite high critical 
temperatures in (Ge,Mn) films\cite{liApl,cho,pinto,tsui}. However, the underlying reasons were not really clear. In Ref.\cite{kang} scanning photoelectron microscopy 
measurements revealed stripe-shaped  Mn rich microstructures which were believed to be the origin of ferromagnetism in Ge$_{1-x}$Mn$_x$. More recent experimental studies 
have revealed self-organized Mn rich nanocolumns formation in Ge$_{1-x}$Mn$_x$, which gave rise to a very high T$_C$ ($\ge$400 K)\cite{jamet} for $x$=0.06. 
Magnetotransport measurements, in this case, have also shown a large anomalous Hall effect up to room temperature. The spinodal decomposition 
(alternating regions of low and high concentration of magnetic impurities) was suggested to be the reason for the high temperature ferromagnetism in this case. 
Similar nanometer-sized clusters, with increased Mn content compared to the surrounding matrix, were also detected by transmission electron microscopy (TEM) analysis in 
Ge$_{0.95}$Mn$_{0.05}$\cite{bougeard}. In recent experimental studies on (Zn,Co)O\cite{jedrecy}, the authors claimed the existence of two types of nanosized 
ferromagnetic Co clusters. The first were spherical with diameters of about 5 nm leading to critical temperatures of $\sim$100 K and the others were columnar about 
4 nm wide, with a maximum height of 60 nm, leading to significantly larger critical temperatures of $\sim$300 K. These results were confirmed by high-resolution 
transmission electron microscopy (HRTEM). Hence this kind of anisotropic nanoscale inhomogeneity can lead to interesting magnetic and transport properties. Inspite of 
the existence of several experimental studies, the effect of impurity clustering on magnetism in DMSs and DMOs has been weakly studied on the theoretical front. \textit {Ab 
initio} based studies for these type of inhomogeneous disordered systems are difficult due to the large size of supercells required and no standard methods have been 
proposed as yet. In Ref.\cite{satojjap} the authors have simulated the spinodal decomposition in DMS by using Monte Carlo methods and they have predicted an above-room-temperature T$_C$ for
the spinodal phase in (Ga,Mn)As and (Ga,Mn)N, calculated from the ``standard'' random-phase approximation (RPA). Here ``standard'' means that the crucial self consistency was 
not implemented in the RPA calculations. However these high T$_C$s were found for samples containing a relatively high concentration of Mn, above 20\%, far from the 
dilute regime. On the other hand in the dilute case, for approximately 5\% of Mn, the authors have found a suppression of the critical temperatures in the presence of 
spinodal decomposition phases. Note also that the calculations were limited to small system sizes compared to the typical size of the inhomogeneities and the average 
was done over few configurations only (typically 10). In Ref.\cite{rao} the authors have presented density-functional theory (DFT) based calculations of N-doped Mn 
clusters, and have given a hypothesis that a high Curie temperature detected in some of the GaMnN samples is a result of the formation of small Mn clusters carrying giant 
magnetic moments. The large variation in Curie 
temperatures could be attributed to the formation of N induced Mn clusters of different sizes in samples grown under different conditions. Their analysis suggests the 
importance of the growth mechanism in these kind of materials. Similar density-functional calculations on the effect of microscopic Mn clustering on the Curie
temperatures of (Ga,Mn)N were also reported in Ref.\cite{hynninen}. However the T$_C$'s were calculated from the mean-field approximation, which is already known to 
overestimate the Curie temperatures in homogeneously diluted semiconductors.

In this article, we present a generalized and comprehensive study of the effect of nanoscale inhomogeneities on the Curie temperatures in 
diluted magnetic systems. The calculations are performed on very large systems (finite size analysis is provided) and a systematic sampling  is done over several 
hundreds of disorder configurations.  In contrast to previous studies, we report giant effects on the T$_C$ in dilute materials.  In some particular cases the T$_C$ can be 
enhanced by up to 1600\% compared to that of the homogeneously diluted system. There are several factors that lead to these effects, such as 
the concentration of inhomogeneities in the system, the size of the inhomogeneities, the concentration of magnetic impurities inside the 
inhomogeneities and also the range of the exchange interactions between the impurities. In the following we shall see how these physical parameters play
an important role and affect the critical temperatures.

\begin{figure}[t]\centerline
{\includegraphics[width=5.4in,angle=0]{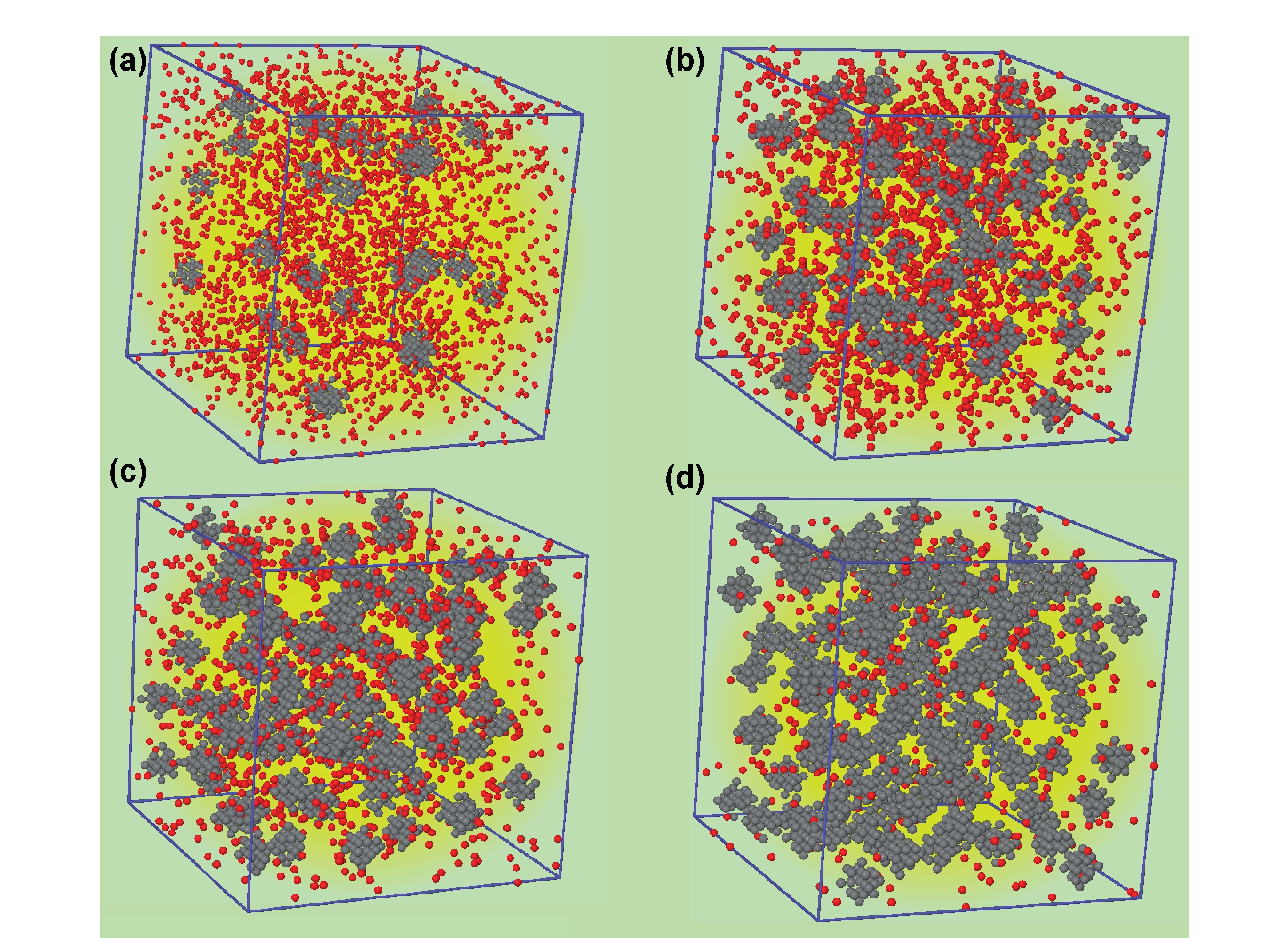}}
\caption{(Color online)Snapshots corresponding to four different concentrations of nanospheres $x_{ns}$ (a) 0.02, (b) 0.04, (c) 0.06 and (d) 0.08. The grey (red) atoms denote the impurities
 inside (outside) the nanospheres. Here L=36, r$_0$=2$a$ ($a$ is the lattice spacing) and $x_{in}$=0.8.
} 
\label{fig1}
\end{figure} 

\section{MODEL AND METHOD}
For simplicity we have assumed here a simple cubic crystalline structure and the conclusions that will be drawn will be general. The sizes vary from L=32 to L=44. The 
inhomogeneities considered here are of spherical shape of radii r$_0$. For the sake of clarity and to avoid additional parameters, in our calculations the total 
concentration of impurities in the whole system is fixed to $x$=0.07. In the following we denote the concentration of nanospheres by $x_{ns}$=N$_{S}$/N, where 
N$_{S}$ is the total number of sites included in all the nanospheres and  N=L$^3$ is the total number of sites. The concentration of impurities inside each nanosphere is 
defined by $x_{in}$. We denote  the total number of impurities and the number of impurities inside the nanospheres by N$^{tot}_{imp}$ and 
N$^{in}_{imp}$ (=$x_{in}$N$_{S}$) respectively. We choose the nanospheres in such a manner so as to restrict their overlap with each other. 

 In Fig.~\ref{fig1} four typical  random configurations corresponding to four different concentration of nanospheres $x_{ns}$ (0.02, 0.04, 0.06 and 0.08) are depicted. As we 
increase the number of nanospheres in the system $x_{ns}$ increases and consequently the concentration outside decreases, since the total concentration ($x$) is fixed. Now 
to evaluate the T$_C$, the effective diluted Heisenberg Hamiltonian H$_{Heis}$=-$\sum_{i,j}$ J$_{ij}$ ${\bf S}_{i}\cdot{\bf S}_{j}$, is treated within the SC-LRPA theory. 
The self-consistent local RPA is a semi-analytical approach based on finite temperature Green's functions. It is essentially an extension of the standard RPA to the 
case of disordered systems. Here the thermal fluctuations are treated within the RPA and the disorder is treated exactly without any approximations. The Curie temperature 
of a system containing $N_{imp}$ localized spins is obtained from the expression  

\begin{equation}
\label{eq.1}
k_{B}T_{C}=\frac{2}{3}S(S+1) \frac{1}{N_{imp}}\sum_{i} \frac{1}{F_{i}}
\end{equation}
where
\begin{equation}
\label{eq.2}
F_{i}=-\frac{1}{2\pi\lambda_{i}}\int_{-\infty}^{\infty}\frac{\Im G_{ii}(E)}{E}dE
\end{equation}
 
We define the retarded Green's function as $G_{ij}(\omega)$=$\int_{-\infty}^{\infty}G_{ij}(t)e^{i\omega t}dt$=$\langle\langle S_i^+;S_j^- \rangle\rangle$. The set of
 parameters $\lambda_{i}$=lim$_{T\rightarrow T_C}$$\langle{S_i^z}\rangle/m$, where $m$ is the average magnetization, are calculated self-consistently (more details 
can be found in Ref\cite{satormp,georges1}). The T$_C$ is calculated for each random configuration and then averaged over a few hundred configurations of disorder. The 
accuracy and reliability of the SC-LRPA to treat disorder and/or dilution has been demonstrated several times in the past\cite{satormp,richard1,akash}.

The exchange couplings in a DMS, as found from \textit{ab initio} based calculations, are relatively short range in nature and almost exponentially 
decaying\cite{bergqvist,satoprb}. Thus in the present study we have assumed generalized couplings of the form  J$_{ij}$=J$_{0}$exp$(-\mid \textbf r \mid/\lambda)$, where 
\textbf r=\textbf r$_i$-\textbf r$_j$ and $\lambda$ is the damping parameter. In (Ga,Mn)As, for about 5\% Mn a fit of the \textit{ab initio} magnetic couplings provides a value 
of $\lambda$ of the order of a/2. Note that in the case of (Ga,Mn)N the \textit{ab initio} couplings are of even shorter range. Thus we focus here on two particular cases, 
$\lambda$=$a$ and $\lambda$=$a/2$, where $a$ is the lattice spacing. Although these length scales are comparable, in the presence of inhomogeneities 
the effects on the critical temperatures will be very drastic.  In order to measure directly the effects of nanoscale inhomogeneities, the averaged Curie temperatures 
$\langle T_C^{inh} \rangle$ are scaled with respect to the averaged Curie temperatures of the homogeneously diluted system $\langle T_C^{hom}\rangle$ for $x$=0.07, their 
ratio is denoted by $\langle$R$_C$$\rangle$. The averaged Curie temperatures $\langle T_C^{hom}\rangle$ for the homogeneous systems are found to be  0.9 J$_{0}$ and 
0.05 J$_{0}$, for $\lambda$=$a$ and $a/2$ respectively, for $x$=0.07.

In Table~(\ref{tbl1}) we provide the averaged Curie temperatures for 80\%, 70\%, 60\% and 40\% homogeneously distributed magnetic impurities scaled with respect to that of the 7\%
homogeneous case, for $\lambda$=$a$ and $a/2$. The ratio is denoted by $\langle$R$^{hom}$$\rangle$. These values will be relevant in the discussions 
to follow, where we consider these types of concentrations inside the nanospheres.

\begin{table}
  \caption{The ratio, $\langle$R$^{hom}$$\rangle$, of the homogeneous Curie temperatures for different $x$ to that of $x$=0.07, for $\lambda$=$a$ and $a/2$}.
  \label{tbl1}
  \begin{ruledtabular}	
  \begin{tabular}{ccc}
    \hline
    \textit x & $\langle$R$^{hom}$$\rangle$ ($\lambda$=$a$) & $\langle$R$^{hom}$$\rangle$ ($\lambda$=$a/2$)  \\
    \hline
    0.8 & 9.7 & 22 \\
    0.7 & 8.9 & 20 \\
    0.6 & 7.9 & 17 \\
    0.4 & 5.2 & 11 \\
    \hline
  \end{tabular}
  \end{ruledtabular}
\end{table}

\begin{figure}[htbp]
\centering
\subfigure{
\includegraphics[width=3.4in,angle=0]{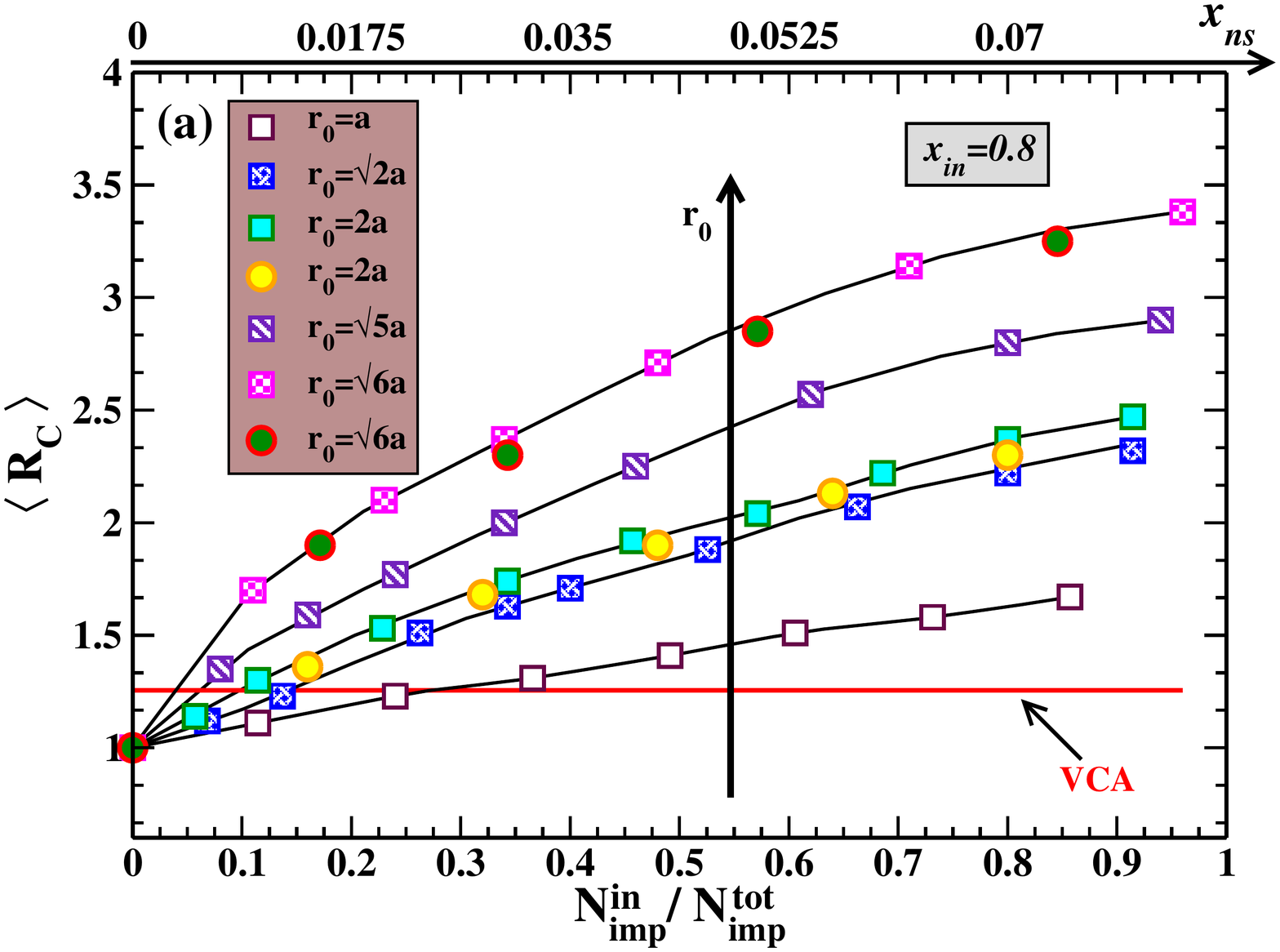}
\label{fig2a}
}
\subfigure{
\includegraphics[width=3.4in,angle=0]{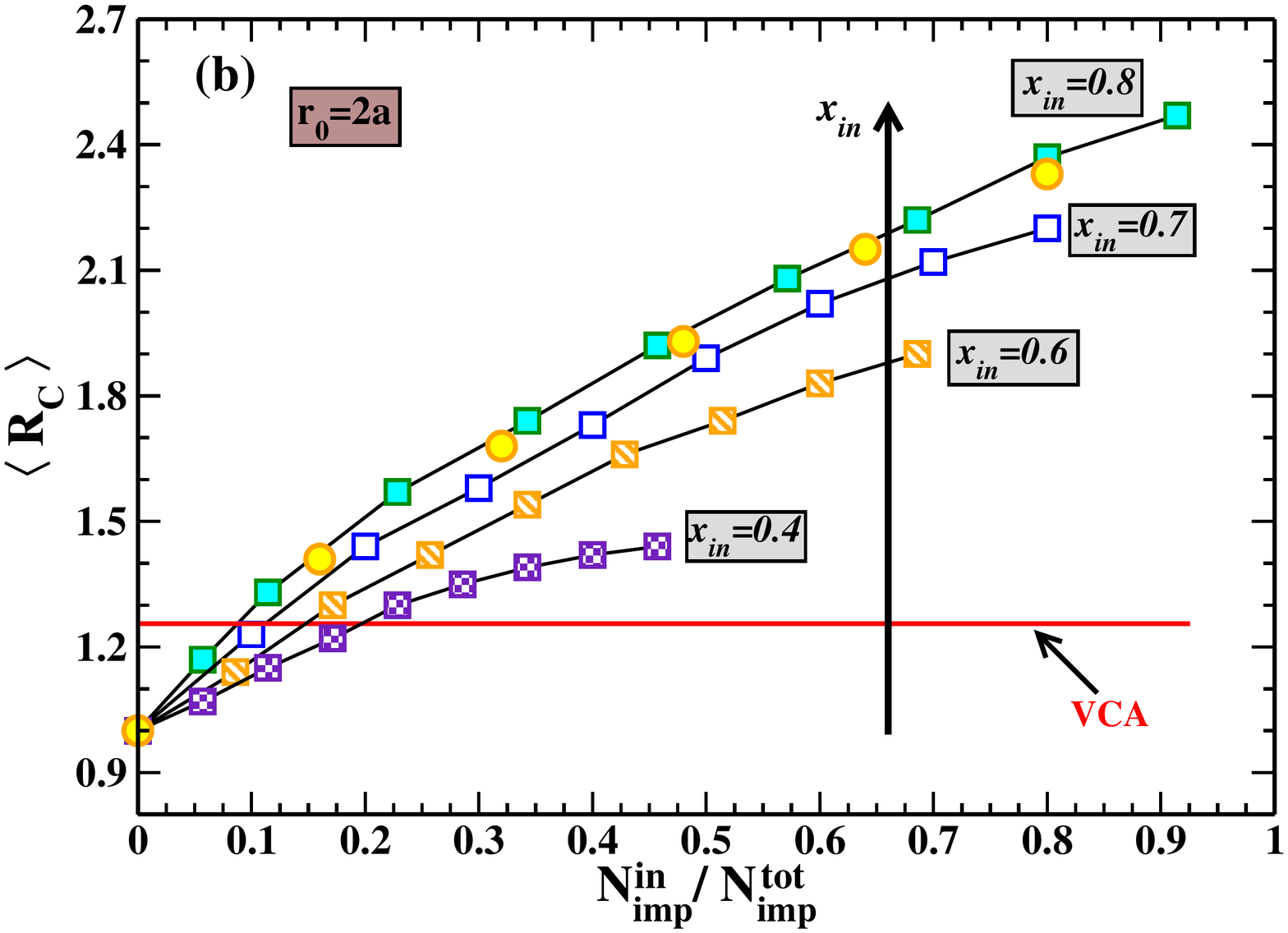}
\label{fig2b}
}
\label{fig2}
\caption[Optional caption for list of figures]{
(Color online)$\langle$R$_C$$\rangle$=$\frac{\langle T_C^{inh} \rangle}{\langle T_C^{hom}\rangle}$ as a function of N$^{in}_{imp}$/N$^{tot}_{imp}$=$\frac{x_{in}}{x}$$x_{ns}$ 
for $\lambda$=$a$. (a) Results for a fixed concentration inside the nanospheres ($x_{in}$=0.8) and different radii r$_0$. The upper $x$ axis represents the values 
of $x_{ns}$ corresponding to $x_{in}$=0.8. The long black arrow indicates the direction of increasing r$_0$. (b) Results for a fixed radius (r$_0$=2$a$) and different concentration 
inside the nanospheres. The long black arrow indicates the direction of increasing $x_{in}$. The solid red line indicates the T$_C^{VCA}$ scaled with respect to 
$\langle T_C^{hom}\rangle$. In the figures, squares correspond to L=32 and circles to L=36.
}
\end{figure} 

\section{RESULTS AND DISCUSSION}
Figure~\ref{fig2a} shows $\langle$R$_C$$\rangle$ as a function of N$^{in}_{imp}$/N$^{tot}_{imp}$ corresponding to the case of $\lambda$=$a$. The concentration inside the nanospheres is fixed 
at $x_{in}$=0.8 and the T$_C$ is calculated for spheres of different radii. For this concentration inside $x_{in}$, each nanosphere contains 5, 15, 26, 45, and 64 impurities
 for r$_0$=$a$, $\sqrt{2}a$, 2$a$, $\sqrt{5}a$, and $\sqrt{6}a$ respectively. N$^{in}_{imp}$/N$^{tot}_{imp}$=0 corresponds to the homogeneously diluted 
case (absence of inhomogeneities). We observe a clear increase in the critical temperatures with increasing fraction of impurities inside the nanospheres as well as with the 
nanospheres' size. For about 80\% of the total impurities inside the nanospheres, T$_C$ is enhanced by up to 150\% for the smallest nanospheres with r$_0$=$a$, and by 
almost 350\% for the ones of radius r$_0$=$\sqrt{6}a$, which is rather significant. This increase for r$_0$=$\sqrt{6}a$ is more than one-third of that found for the 80\%
homogeneously distributed case (Table~\ref{tbl1}). Thus the clustering of magnetic impurities does lead to a considerable 
increase of the critical temperatures due to the strong interactions within the nanospheres. The other important point to take note of is the T$_C$ obtained from the mean 
field virtual crystal approximation (VCA), T$_C^{VCA}$=$\frac{2}{3}x\sum_{i}$n$_{i}$J$_{i}$, where n$_{i}$ is the number of atoms in the $i$-th shell. It is well known 
that the VCA overestimates the true critical temperatures,
often very strongly. However, the present results show that in the presence of inhomogeneities the VCA value can no longer serve as an upper bound. Indeed, as can be 
seen here, for a relatively small concentration of nanospheres ($x_{ns}$$\sim$0.2) the VCA value is already exceeded and for higher 
density of nanospheres the VCA actually strongly underestimates the critical temperatures in these systems.

Let us now focus on the case where the nanospheres are of fixed radius (r$_0$=$2a$) and the concentration inside the nanospheres vary (Fig.~\ref{fig2b}).{$\langle$R$_C$$\rangle$ is 
plotted as a function of N$^{in}_{imp}$/N$^{tot}_{imp}$ for different $x_{in}$. The curves show an overall monotonous increase with increasing concentration of nanospheres.
 However, the enhancement of the critical temperatures is also controlled by the concentration of impurities inside the nanospheres. Decreasing the concentration inside the 
nanospheres effectively means reducing the number of impurities inside a cluster of the same size, and thus reducing inter nanosphere interactions. This could explain the 
relatively small increase in the T$_C$ values with decreasing $x_{in}$. However, as will be seen in the following, the variation of the critical temperatures is more 
complex than this simple picture.  Thus we find that not only the relative number of impurities inside the nanospheres but also the concentration inside the nanospheres 
have a drastic effect on the critical temperatures in these systems.

\begin{figure}[htbp]
\centering
\subfigure{
\includegraphics[width=3.4in,angle=0]{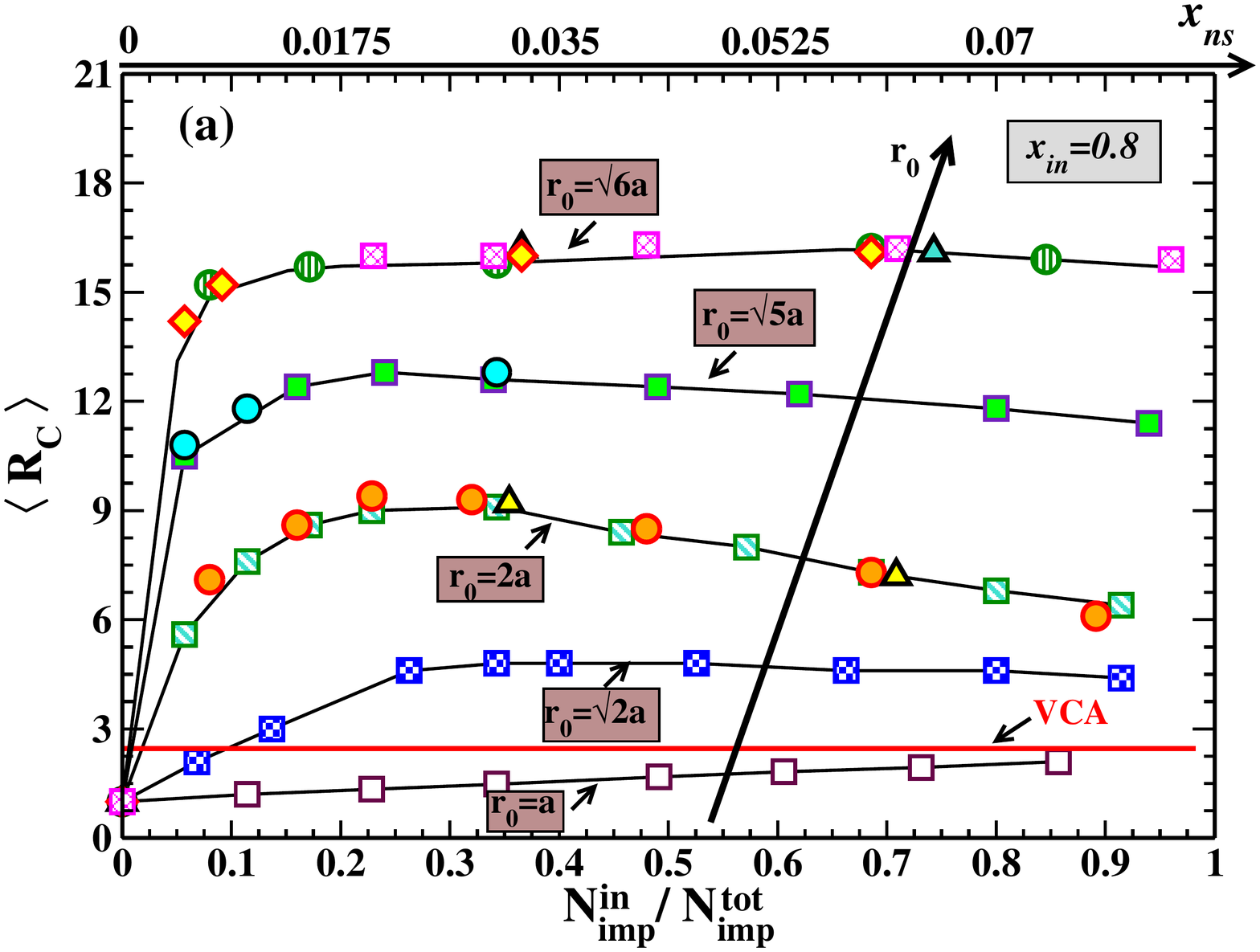}
\label{fig3a}
}
\subfigure{
\includegraphics[width=3.4in,angle=0]{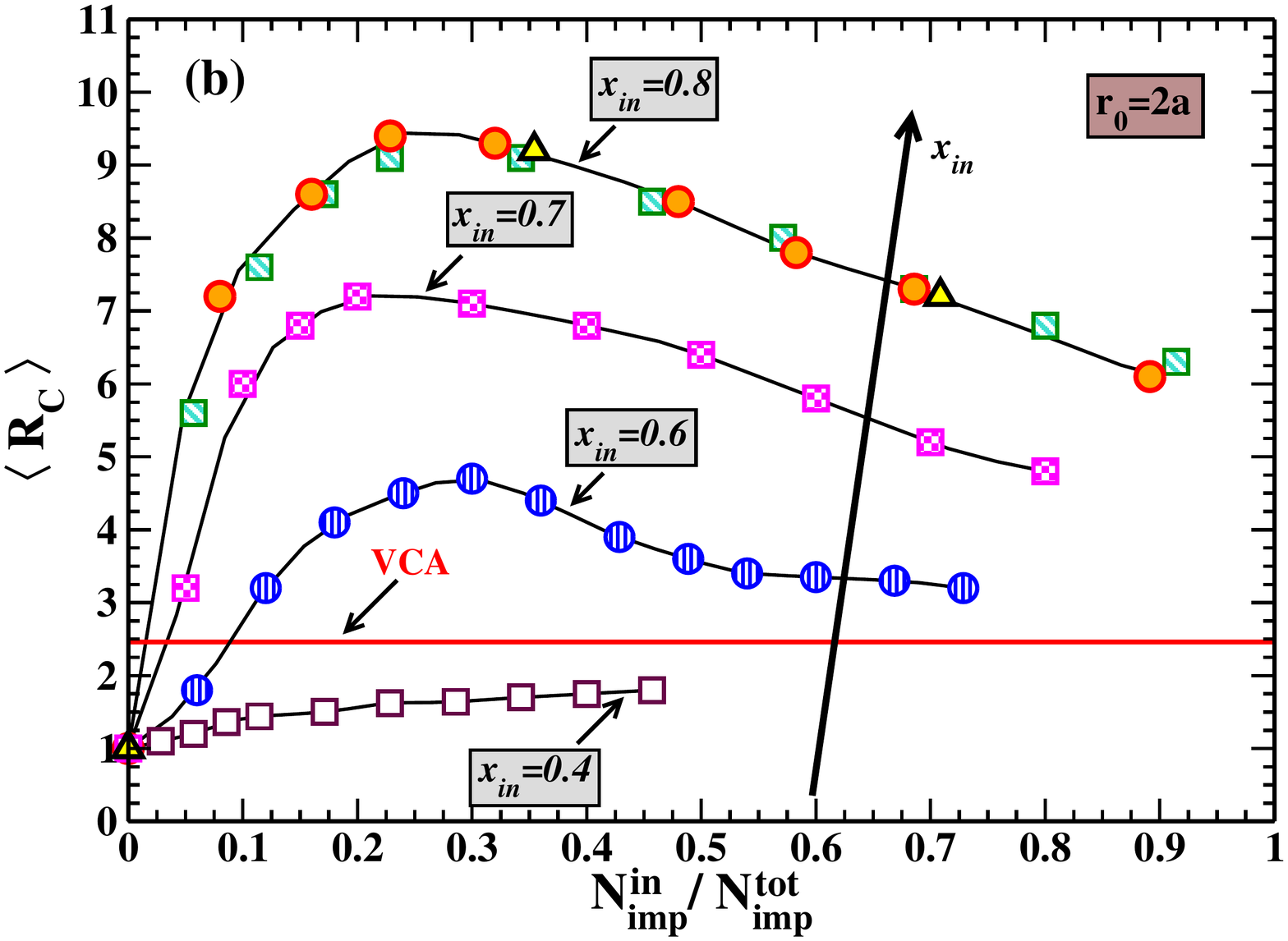}
\label{fig3b}
}
\label{fig3}
\caption[Optional caption for list of figures]{
(Color online)$\langle$R$_C$$\rangle$=$\frac{\langle T_C^{inh} \rangle}{\langle T_C^{hom}\rangle}$ as a function of N$^{in}_{imp}$/N$^{tot}_{imp}$=$\frac{x_{in}}{x}$$x_{ns}$
 corresponding to $\lambda$=$a/2$. (a) Results for a fixed concentration inside the nanospheres ($x_{in}$=0.8) and different radii r$_0$. The upper $x$ axis 
represents the values of $x_{ns}$ corresponding to $x_{in}$=0.8. The long black arrow indicates the direction of increasing r$_0$. (b)Results for a fixed radius (r$_0$=2$a$) and different concentration inside the nanospheres. 
The long black arrow indicates the direction of increasing $x_{in}$. The solid red line indicates the T$_C^{VCA}$ scaled with respect to $\langle T_C^{hom}\rangle$. In the figures, squares correspond to L=32, circles 
to L=36, triangles to L=40, and diamonds to L=44.
}
\end{figure}

Now we move to the case of the shorter-ranged couplings, $\lambda$=$a/2$, which will appear even more interesting and which lead to unexpected effects. Figure~\ref{fig3a} shows the 
$\langle$R$_C$$\rangle$ as a function of N$^{in}_{imp}$/N$^{tot}_{imp}$ for a fixed $x_{in}$=0.8. T$_C$ is calculated for nanospheres of different radii (r$_0$=$a$, $\sqrt{2}a$, 2$a$, 
$\sqrt{5}a$ and $\sqrt{6}a$). We have considered system sizes varying from L=32 to L=44 to check for the finite-size effects. The L=44 systems typically contain 
$\sim$6000 impurities. In contrast to the case of $\lambda$=$a$ discussed above, the variation of T$_C$ with N$^{in}_{imp}$/N$^{tot}_{imp}$ is not monotonous anymore. Here we see a colossal 
effect of the size of the nanospheres on the T$_C$. For the smallest nanospheres (r$_0$=$a$) there is hardly any noticeable effect, with the critical temperatures 
remaining close to that of the homogeneous case. Now as we increase the radius of the nanospheres for a given concentration of nanospheres, there is a sharp and strong increase in the T$_C$ 
values. As can be seen, even for a reasonably small concentration of nanospheres ($x_{ns}$$\sim$0.2) we obtain a remarkable 
jump of almost 900\% for r$_0$=2$a$ and even 1600\% for r$_0$=$\sqrt{6}a$, compared to that of the homogeneous case. This gigantic increase in the presence of nanospheres 
with r$_0$=$\sqrt{6}a$ is more than 70\% when compared to the T$_C$ of the 80\% homogeneous case (Table~\ref{tbl1}), which is rather extraordinary.} This implies that in materials
 such as (Ga,Mn)N, where the exchange interactions are 
really short ranged, it would be possible to reach T$_C$$\ge$500 K (as T$_C$ for homogeneously diluted Ga$_{1-x}$Mn$_{x}$N is 40 K for $x$=0.07\cite{bergqvist,georges1,
richard1}) by inducting nanoscale inhomogeneities. The  presence of such nanoclusters may also explain the very high T$_C$s observed in Ga$_{1-x}$Mn$_{x}$N by some 
experimental groups\cite{sonoda}. It should be of great  interest to analyse experimentally the effect of such nanoclusters on the critical temperatures in these kind of 
materials. Here again the mean field VCA is found to strongly underestimate the T$_C$ for most cases. This is expected since the mean field VCA treatment is unable to
 capture all the relevant physical effects in both homogeneously disordered as well as inhomogeneous systems. Thus it becomes clear that in systems with relatively short-
ranged couplings the size of the inhomogeneities plays a very important role in controlling the critical temperatures.  The nonmonotonous behavior observed here implies 
that several physical parameters are in competition (length scales and relevant couplings). Thus, we cannot explain this variation by assuming the inhomogeneities to behave 
as ``super-spins'' only. 

In Fig.~\ref{fig3b} we consider the case of nanospheres of fixed radii r$_0$=2$a$, which is particularly interesting. $\langle$R$_C$$\rangle$ is shown as a function of 
N$^{in}_{imp}$/N$^{tot}_{imp}$ for different $x_{in}$ (0.8, 0.7, 0.6 and 0.4). For a fixed $x_{in}$, we observe a gradual increase in the critical temperatures with increasing concentration of
nanospheres, and then it decreases as $x_{ns}$ increases further. In contrast to the case of $\lambda$=$a$, there is a clear maximum in the 
T$_C$ around N$^{in}_{imp}$/N$^{tot}_{imp}$$\sim$0.2 for $x_{in}$=80\% and 70\%. For this value of N$^{in}_{imp}$/N$^{tot}_{imp}$, as we increase the concentration inside the 
nanospheres we observe a huge jump in the critical temperatures, from a small increase for $x_{in}$=40\% to almost  900\% for $x_{in}$=80\%, compared to that of the 
homogeneous case. It should be noted that for N$^{in}_{imp}$/N$^{tot}_{imp}$=0.9 and $x_{in}$=80\% the increase is reduced to about 600\%, which is 
still considerably large. However for $x_{in}$=40\% we hardly obtain any significant increase compared to that of the homogeneous case.
 Hence in this case the concentration inside the nanospheres is found to have a crucial effect on determining the critical temperatures of the system. A careful 
statistical analysis reveals that the case of r$_0$=2$a$ for $\lambda$=$a/2$ is particularly intriguing. As will be seen, the analysis of the T$_C$ distributions
exhibits interesting features. In the following we provide a more detailed study for this particular case and try to analyse the reasons for the origin of this kind of 
behavior.

\begin{figure}[htbp]
\centerline
{\includegraphics[width=5.4in,angle=0]{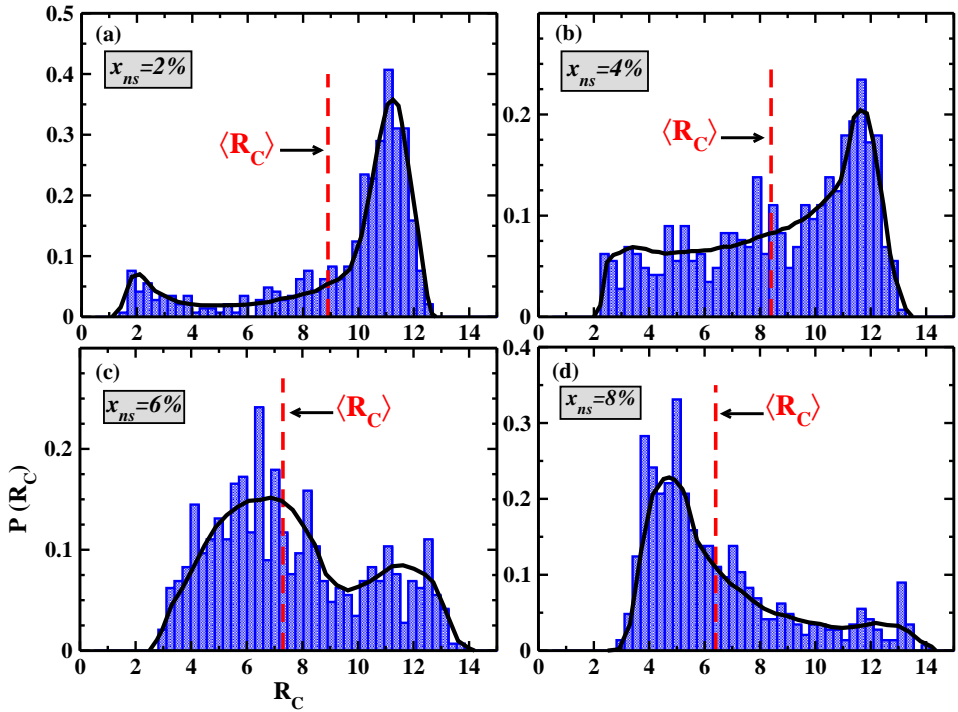}}
\caption{(Color online)Normalized R$_C$ distributions for four different $x_{ns}$: (a) 0.02, (b) 0.04, (c) 0.06, and (d) 0.08 corresponding to $\lambda$=$a/2$. Here r$_0$=2$a$,
$x_{in}$=0.8 and L=32. The red dashed lines indicate the $\langle$R$_C$$\rangle$ values, which we have shown in Fig.\ref{fig3b}. The solid black lines are a guide to the eye.
} 
\label{fig4}
\end{figure}

In Fig.~\ref{fig4} we show the normalized R$_C$ distributions corresponding to the case of r$_0$=2$a$, $x_{in}$=0.8 and $\lambda$=$a/2$. The distributions are obtained 
using a sampling over a few hundred configurations of disorder ($\sim$600). As can be seen from the figure, we obtain very interesting wide distributions for the different 
concentrations of nanospheres ($x_{ns}$=0.02, 0.04, 0.06 and 0.08). For $x_{ns}$=0.02 we observe a kind of bimodal distribution: one peak with a large weight at high 
T$_C$ (T$_C^{high}$$\sim$11${\langle T_C^{hom}\rangle}$) values and another one at lower T$_C$ (T$_C^{low}$$\sim$2${\langle T_C^{hom}\rangle}$) with a much smaller weight.
 When increasing $x_{ns}$ to 
0.04, the width of the distribution is almost unaffected, but we notice a clear transfer of weight from the high T$_C$ values to the lower one.  By further increasing 
$x_{ns}$ to 0.06, the transfer of weight increases further: the low-T$_C$ region has a significantly higher weight. Finally for relatively high $x_{ns}$ ($\sim$0.08), the 
weight is now concentrated around the lower T$_C$ values and the distribution exhibits a tail-like structure at higher critical temperatures. This transfer of weight is the 
reason for the maximum in the T$_C$ observed in Fig.~\ref{fig3a}. The origin of this kind of distribution is not very clear at first. However the analysis of the 
configurations reveals an interesting feature. We have considered two different kind of configurations. The first set of configurations of nanospheres corresponds to the 
situation where the distance between the nanospheres is restricted to small separations. The second kind corresponds to large separations between the nanospheres. First, it is 
found that in both cases the  distribution of T$_C$ is relatively narrow and unimodal. However, in the first case the T$_C$ distribution is centered around T$_C^{low}$, 
whereas in the second case it is centered around  T$_C^{high}$. It is surprising and counter-intuitive that the largest T$_C$'s are obtained from the configurations 
where the inter nanosphere couplings are weaker. This is a clear indication that several length scales are competing. Now the nature of distributions shown in Fig.~\ref{fig4} 
can be explained as follows. In the case of low concentration of nanospheres (Fig.~\ref{fig4}(a)) the probability of finding the  nanospheres relatively far apart from each other is 
relatively high, and conversely the probability of finding them close to each other is relatively small. Thus this  leads, in the distribution, to a significant weight 
around the high T$_C$ values. As we gradually increase $x_{ns}$, the probability of finding configurations with the nanospheres at relatively large separation decreases, while 
the probability corresponding to small separation increases. As a consequence the weight in the distribution around T$_C^{high}$ decreases and that corresponding to the low T$_C$ 
increases, as observed in Figs.~\ref{fig4}(b) and ~\ref{fig4}(c). Finally for the largest $x_{ns}$($\sim$0.08) the weight is mainly concentrated around the low-T$_C$ region (Fig.~\ref{fig4}(d)). Interestingly, 
this kind of behavior is not observed in the case of $\lambda$=$a$, for nanospheres of radii varying from r$_0$=$a$ to $\sqrt{6}a$.  For this case (longer range) 
the distribution of the critical temperatures is always narrow and unimodal, thus all the configurations (nanospheres far apart or close to each other) lead to similar values of the
 critical temperature. This confirms the idea that several length scales and typical couplings compete to give rise to this rich and new physics.

Let us now discuss some experimental consequences. We have shown that in systems with effective short-ranged exchange interactions it is possible to obtain two different 
critical temperatures depending on the size and concentration of the inhomogeneities and also on the typical separation between them. For example, in the case of 
$\lambda$=$a/2$ for nanospheres of radii r$_0$=2$a$, $x_{ns}$=0.02 and $x_{in}$=0.8, the T$_C^{high}$ value is almost \textit{five} times that of the T$_C^{low}$ 
value. This could explain the wide range of T$_C$ values observed  experimentally for materials such as (Ga,Mn)N\cite{overberg,theodoropoulou,reed,thaler,sonoda} and the 
apparent dissension between theoretical predictions and experimental observations for these kinds of materials. In this context, it should be noted that 
Li \textit {et al.}\cite{liPrb} proposed two different ordering  temperatures in Ge$_{1-x}$Mn$_x$, T$_C$ and T$_C^{*}$ with T$_C$$\ll$T$_C^{*}$. The higher critical 
temperature T$_C^{*}$ is associated with the ferromagnetic ordering temperature within isolated spin clusters and the onset of global ferromagnetism only occurs at 
T$_C$. For $x$=0.05 the values of T$_C$ and T$_C^{*}$ were found to be 12 and 112 K, respectively. However, detailed experimental studies in this direction would help to 
fully confirm this picture.

\section{CONCLUSION}
In conclusion, we have presented a detailed study of the effect of nanoscale inhomogeneities on the critical temperatures in diluted magnetic systems. We have shown that 
for materials with effective short ranged exchange interactions it is indeed possible to go beyond room temperature ferromagnetism by inducting nanoscale clusters of 
magnetic impurities. A gigantic increase in the critical temperatures of up to 1600\%, compared to that of the homogeneously diluted case, is obtained in certain 
cases. We also provide a plausible
explanation for the wide variation of the T$_C$'s, observed experimentally, in some materials such as (Ga,Mn)N. A meticulous study revealed that the relative separation 
between the inhomogeneities can play a decisive role in controlling the Curie temperatures. In some cases uniform distribution of nanospheres 
 is found to favor very high critical temperatures. This fact could be further
 corroborated by detailed experimental studies. If, by controlling the growth conditions, the formation of the nanoscale inhomogeneities can be manipulated, it will open up
 the possibility of studying these disordered inhomogeneous systems in more detail. We believe that our study will pave the way for a better understanding of the origin and
 control of high-temperature ferromagnetism in dilute magnetic systems, which can serve as building blocks for potential future spintronic devices.

\acknowledgments
We acknowledge Denis Feinberg, Claudine Lacroix, Arnaud Ralko and Paul Wenk for valuable discussions and insightful comments. S.K. gratefully acknowledges 
support by the WCU program (R31-2008-000-10059-0) AMS. A.C. would like to thank the Nanosciences Fondation for financial support.


\begin{thebibliography}{99}

\bibitem{satormp} K. Sato, L. Bergqvist, J. Kudrnovsk\'y, P. H. Dederichs, O. Eriksson, I.Turek, B. Sanyal,
G. Bouzerar, H. Katayama-Yoshida, V. A. Dinh, T. Fukushima,H. Kizaki and R. Zeller, Rev. Mod. Phys. \textbf{82,} 1633 (2010).

\bibitem{jungwirth} T. Jungwirth, J. Sinova, J. Masek, J. Kucera and A. H. MacDonald, Rev. Mod. Phys. \textbf{78,} 809 (2006).

\bibitem{timm} C. Timm, J. Phys. Condens. Matter \textbf{15,} R1865 (2003); 
C. Timm, F. Sch\"afer and F. von Oppen,  Phys. Rev. Lett. \textbf{89,} 137201 (2002).

\bibitem{sato-yoshida} K. Sato and H. Katayama-Yoshida, Semicond. Sci. Technol \textbf{17,} 367 (2002).

\bibitem{fukumura} T. Fukumura, H. Toyosaki and Y. Yamada, Semicond. Sci. Technol \textbf{20,} S103 (2005).

\bibitem{chambers} S. A. Chambers, T. C. Droubay, C. M. Wang, K. M. Rosso, S. M. Heald, D. A. Schwartz, K. R. Kittilstved and D. R. Gamelin, 
 Mater. Today \textbf{9,} 28 (2006).

\bibitem{overberg} M.E. Overberg, C. R. Abernathy, S. J. Pearton, N. A. Theodoropoulou, K. T. McCarthy and A. F. Hebard, Appl. Phys. Lett. \textbf{79,} 1312 (2001).

\bibitem{theodoropoulou} N. Theodoropoulou, A. F. Hebard, M. E. Overberg, C. R. Abernathy, S. J. Pearton, S. N. G. Chu,  
and R. G. Wilson, Appl. Phys. Lett. \textbf{78,} 3475 (2001).

\bibitem{reed} M. L. Reed, N. A. El-Masry, H. H. Stadelmaier, M. K. Ritums, M. J. Reed, C. A. Parker, J. C. Roberts and S. M. Bedair,
 Appl. Phys. Lett. \textbf{79,} 3473 (2001).

\bibitem{thaler} G. T. Thaler, M. E. Overberg, B. Gila, R. Frazier, C. R. Abernathy, S. J. Pearton, J. S. Lee, S. Y. Lee, Y. D. Park,
Z. G. Khim, J. Kim and F. Ren, Appl. Phys. Lett. \textbf{80,} 3964 (2002).

\bibitem{sonoda} S. Sonoda, S. Shimizu, T. Sasaki, Y. Yamamoto and H. Hori, J. Cryst. Growth \textbf{237-239,} 1358 (2002).

\bibitem{richard1} R. Bouzerar and G. Bouzerar, Europhys. Lett. \textbf{92,} 47006 (2010).

\bibitem{georges1} G. Bouzerar, T. Ziman and J. Kudrnovsk\'y, Europhys. Lett. \textbf{69,} 812 (2005).

\bibitem{bergqvist} L. Bergqvist, O. Eriksson, J. Kudrnovsk\'y, V. Drchal, P. Korzhavyi and I. Turek, Phys. Rev. Lett. \textbf{93,} 137202 (2004).

\bibitem{satoprb} K. Sato, W. Schweika, P. H. Dederichs and H. Katayama-Yoshida, Phys. Rev. B \textbf{70,} 201202(R) (2004).

\bibitem{park} Y.D. Park, A. T. Hanbicki, S. C. Erwin, C. S. Hellberg, J. M. Sullivan, J. E. Mattson, T. F. Ambrose, A. Wilson, G. Spanos 
and B. T. Jonker, Science \textbf{295,} 651 (2002).

\bibitem{liApl} A. P. Li, J. Shen, J. R. Thompson and H. H. Weitering, Appl. Phys. Lett. \textbf{86,} 152507 (2005).

\bibitem{cho} S. Cho, S. Choi, S. C. Hong, Y. Kim, J. B. Ketterson, B. J. Kim, Y. C. Kim and J. H. Jung, Phys. Rev. B \textbf{66,} 033303 (2002).

\bibitem{pinto} N. Pinto, L. Morresi, M. Ficcadenti, R. Murri, F. D'Orazio, F. Lucari, L. Boarino and G. Amato, Phys. Rev. B \textbf{72,} 165203 (2005).

\bibitem{tsui} F. Tsui, L. He, L. Ma, A. Tkachuk, Y. S. Chu, K. Nakajima and T. Chikyow, Phys. Rev. Lett. \textbf{91,} 177203 (2003).

\bibitem{kang} J. S. Kang, G. Kim, S. C. Wi, S. S. Lee, S. Choi, S. Cho, S. W. Han, K. H. Kim, H. J. Song, H. J. Shin, A. Sekiyama, S. Kasai, 
S. Suga and B. I. Min, Phys. Rev. Lett. \textbf{94,} 147202 (2005).

\bibitem{jamet} M. Jamet, A. Barski, T. Devillers, V. Poydenot, R. Dujardin, P. Bayle-Guillemaud, J. Rothman, E. Bellet-Amalric, A. Marty, 
J. Cibert, R. Mattana and S. Tatarenko, Nat. Mater. \textbf{5,} 653 (2006).

\bibitem{bougeard} D. Bougeard, S.Ahlers, A. Trampert, N. Sircar and G. Abstreiter, Phys. Rev. Lett. \textbf{97,} 237202 (2006).

\bibitem{jedrecy} N. Jedrecy, H. J. von Bardeleben and D. Demaille, Phys. Rev. B \textbf{80,} 205204 (2009).

\bibitem{satojjap} K. Sato, H. Katayama-Yoshida and P. H. Dederichs, Jpn. J. Appl. Phys. \textbf{44,} L948 (2005).

\bibitem{rao} B. K. Rao and P. Jena, Phys. Rev. Lett. \textbf{89,} 185504 (2002).

\bibitem{hynninen} T. Hynninen, H. Raebiger, J. von Boehm and A. Ayuela, Appl. Phys. Lett. \textbf{88,} 122501 (2006).

\bibitem{akash} A. Chakraborty and G. Bouzerar, Phys. Rev. B \textbf{81,} 172406 (2010).

\bibitem{liPrb} A.P. Li, J.F. Wendelken, J. Shen, L. C. Feldman, J. R. Thompson and H. H. Weitering, Phys. Rev. B \textbf{72,} 195205 (2005).

\end{thebibliography}
\end{document}